\documentclass[aps,prb,floatfix,twocolumn,showpacs,superscriptaddress]{revtex4-1}  
\usepackage[utf8]{inputenc}
\usepackage{amsmath}
\usepackage{physics}
\usepackage{mathtools}
\usepackage{amsfonts}
\usepackage{amssymb}
\usepackage{graphicx}
\usepackage{amsthm}
\usepackage{textcomp}
\usepackage{color}
\usepackage[dvipsnames]{xcolor}
\usepackage{lipsum}

 \usepackage{combelow}

\numberwithin{theorem}{section}

\begin{document}

\title{Emergence of two-fold non-Hermitian spectral topology through synthetic spin engineering}

\author{Ronika Sarkar}
\email{ronikasarkar@iisc.ac.in}
\affiliation{Solid State and Structural Chemistry Unit, Indian Institute of Science, Bangalore 560012, India}
\affiliation{Department of Physics, Indian Institute of Science, Bangalore 560012, India}
\author{Ayan Banerjee}
\affiliation{Solid State and Structural Chemistry Unit, Indian Institute of Science, Bangalore 560012, India}
\author{Awadhesh Narayan}
\email{awadhesh@iisc.ac.in}
\affiliation{Solid State and Structural Chemistry Unit, Indian Institute of Science, Bangalore 560012, India}

\date{\today}

\begin{abstract}
The union of topology and non-Hermiticity has led to the unveiling of many intriguing phenomena. We introduce a synthetic spin-engineered model belonging to symmetry class AI, which is a rare occurrence, and demonstrate the emergence of a \emph{multi-fold} spectral topology. As an example of our proposal, we engineer non-Hermiticity in the paradigmatic Su-Schrieffer-Heeger (SSH) model by introducing a generalized synthetic spin, leading to an emergent two-fold spectral topology that governs the decoupled behaviour of the corresponding non-Hermitian skin effect. As a consequence of the spin choice, our model exhibits a rich phase diagram consisting of distinct topological phases, which we characterize by introducing the notion of \emph{paired} winding numbers, which, in turn, predict the direction of skin localization under open boundaries. We demonstrate that the choice of spin parameters enables control over the directionality of the skin effect, allowing for it to be unilateral or bilateral. Furthermore, we discover non-dispersive flat bands emerging within the inherent SSH model framework, arising from the spin-engineering approach. We also introduce a simplified toy model to capture the underlying physics of the emergent flat bands and direction-selective skin effect. As an illustration of experimental feasibility, we present a topoelectric circuit that faithfully emulates the underlying spin-engineered Hamiltonian, providing a viable platform for realizing our predicted effects. Our findings pave way for the exploration of unconventional spectral topology through spin-designed models.
\end{abstract}
\maketitle

\section{Introduction}

The synergy of non-Hermiticity and topology has ushered in a remarkable wave of activity~\cite{bergholtz2021exceptional,ashida2020non,ding2022non,banerjee2023non}. This has led to a variety of intriguing discoveries, including non-Hermitian spectral topology~\cite{okuma2023non}, non-Hermitian skin effect~\cite{yao2018edge,alvarez2018non,zhang2022review,lin2023topological}, as well as unexpected connections to topics in modern mathematics~\cite{hu2021knots,jaiswal2023characterizing,wang2022amoeba,banerjee2023tropical,hu2023non}. The ramified symmetries in non-Hermitian systems give rise to distinctive topological phases with the notion of ``line" and ``point" gap topology~\cite{kawabata2019symmetry}. The role of non-Hermitian time dynamics in spectral topology has revealed new insights into the interplay between non-Hermiticity and topological features~\cite{lee2019topological}, providing a deeper understanding of phenomena such as spectral singularities, i.e., exceptional points (EPs)~\cite{heiss2004exceptional,ding2016emergence,miri2019exceptional,chen2020revealing,kawabata2019classification}, and the emergence of novel topological phases with exotic properties~\cite{lee2016anomalous,shen2018topological,li2020critical,liang2022dynamic,kawabata2022many,longhi2019probing,kunst2018biorthogonal,weidemann2020topological,yokomizo2019non,song2019non,xue2022non}. Traversing through EPs leads to phase transitions, and unveils unconventional features such as intersecting Riemann sheet topology~\cite{dembowski2004encircling,doppler2016dynamically}, manifesting as Berry phases and fractional topological charges~\cite{zhou2018observation}, which have a range of potential applications~\cite{wiersig2014enhancing,ozdemir2019parity,chen2017exceptional}.

In recent years, synthetic degrees of freedom have emerged as a powerful tool in physics, offering a platform to engineer novel electronic, magnetic, and topological properties \cite{pesin2012spintronics,celi2014synthetic,ozawa2019topological}. The key concept is to bring together appropriate degrees of freedom to emulate additional dimensions enabling novel phenomena \cite{sinha2011trapped,anderson2012synthetic,meier2019exploring}. Striking examples from past years include the realization of a pseudospin Hall effect in acoustic systems \cite{weiner2022synthetic}, observation of a synthetic Hall ladder in a photonic cavity \cite{dutt2020single}, to name a few. Among the various degrees of freedom, synthetic spin and pseudospin have been particularly fruitful in the quest for unconventional physics \cite{cooper2019topological,aydougdu2009pseudospin}.

\begin{figure} [b]
    \centering
    \includegraphics[width=0.45\textwidth]{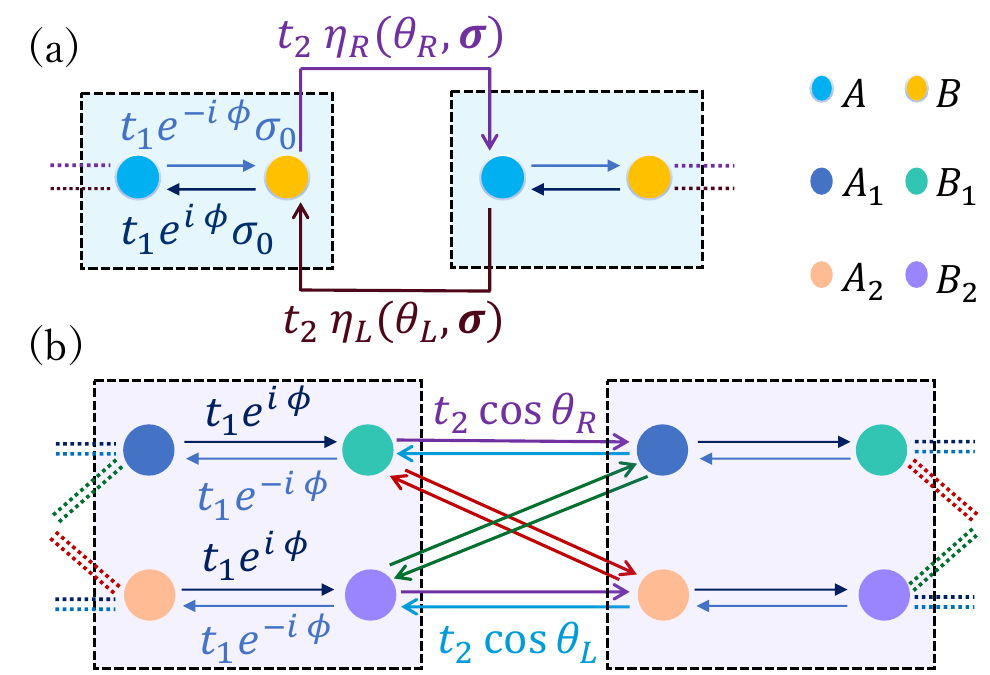}
    \caption{\label{Fig:Model} \textbf{Synthetic spin-engineering in the Su-Schrieffer-Heeger model.} (a) Non-Hermiticity introduced by generalized spin in the SSH model. (b) A simplified schematic of the four-band model for our choice of spin. The non-Hermitian spin associates two degrees of freedom with each unit cell site $A$ and $B$, effectively dissociating them to $A_1, A_2$ and $B_1, B_2$, respectively. Each hopping label corresponds to the same coloured arrow. Additionally, the rightward (leftward) red arrows denote $-i t_2 \sin{\theta_R}$ ($i t_2 \cos{\theta_L}$) and the rightward (leftward) green arrows denote $i t_2 \sin{\theta_R}$ ($-i t_2 \cos{\theta_L}$). A similar approach is directly applicable to other models.}
\end{figure}

In this work, we introduce and develop a rendition of the SSH model, incorporating synthetic-spin engineering, featuring in the rare non-Hermitian symmetry class AI. This approach demonstrates a remarkably rich phase diagram and the emergence of a two-fold spectral topology, which are a direct consequence of the underlying spin structure. We discover a variety of distinct topological phases, which we characterize by introducing \emph{paired} winding numbers. These topological invariants enable the prediction of the skin effect under open boundary conditions, and the selection of spin parameters allows exquisite control over the direction and nature of the skin effect. We show that the spectral topology, as well as the chiral dynamics, are robust to disorder. Notably, the phase diagram features flat bands which are accessible in a wide parameter regime. We further introduce a toy model comprising of coupled Hatano-Nelson chains to understand the underlying theoretical principles of our spin-engineered model. \\

\noindent
\section{Model}

Our proposal is to introduce a synthetic spin to a parent Hermitian model, thus endowing it with additional degrees of freedom. A suitable choice of the spin parameters in turn allows generating non-Hermiticity and the concomitant spectral topology [Fig.~\ref{Fig:Model}(a)]. Originally introduced to study solitons in polyacetylene~\cite{su1979solitons},the paradigmatic SSH model provides a fundamental framework for studying topological systems~\cite{asboth2016schrieffer}. We consider the Hermitian bipartite SSH model with different intra- and inter-unit cell hopping terms. We engineer non-Hermiticity into the model by introducing an artificial spin to the hopping potentials. A general method of engineering a non-Hermitian spin to the inter-unit cell hopping $t_2$ is schematically shown in Fig.~\ref{Fig:Model}(a). The Hamiltonian consists of the intra- and inter-unit cell couplings, i.e., $H=H_\textup{intra}+H_\textup{inter}$. Each component can be expressed as
$
    H_\textup{intra} =\sum_i 
    t_1 e^{-i \phi} \: \sigma_0\: c^\dagger_{A,i}c_{B,i} + h.c.$ and $
    H_\textup{inter} = \sum_i \:[t_2 \eta_R(\theta_R,\vec{\sigma})] \:c^\dagger_{B,i}c_{A,i+1} + [t_2 \eta_L(\theta_L, \vec{\sigma})]\:c^\dagger_{A,i+1}c_{B,i}$, where, $\eta_R(\theta_R,\vec{\sigma}) = \cos{\theta_R}\: \sigma_0 +\sin{\theta_R}\:\sigma_y$ and $\eta_L(\theta_L,\vec{\sigma}) = \cos{\theta_L}\: \sigma_0 +\cos{\theta_L}\:\sigma_y$.

Here, $L$ and $R$ denote leftward and rightward hopping terms, respectively, and $\vec{\sigma}$ are the Pauli matrices ($\sigma_0$ is the identity matrix), which form the generators of the synthetic spin. $\theta_R$ and $\theta_L$ are parameters of this spin (Fig.~\ref{Fig:Model}). 
Starting with the Hermitian SSH Hamiltonian,
$H(k)= \sigma_x \: (t_1 +t_2 \cos{k}) + \sigma_y\: (t_2 \sin{k})$, we incorporate a synthetic spin to each lattice site as shown in Fig.~\ref{Fig:Model}(a). This modifies the SSH model to a \emph{quadripartite} lattice [Fig.~\ref{Fig:Model}(b)]. The Hamiltonian can be written as

\begin{equation}
H=\vec{d}. \:\vec{\Gamma},
\end{equation}

where, $\vec{\Gamma}= (\sigma_x \otimes \sigma_0 , \: \sigma_x \otimes \sigma_y,\: \sigma_y \otimes \sigma_y,\: \sigma_y \otimes \sigma_0)$ and $\vec{d}=(d_1,\: d_2,\: d_3,\: d_4)$, which read,

\begin{equation} 
\begin{split}
    d_1= t_1 + \frac{t_2}{2} \:(e^{ik} \cos{\theta_R} + e^{-ik} \cos{\theta_L)},\\
    d_2= \frac{t_2}{2} \:(e^{ik} \sin{\theta_R} + e^{-ik} \cos{\theta_L)},\\
    d_3= \frac{t_2}{2 i} \:(e^{ik} \sin{\theta_R} - e^{-ik} \cos{\theta_L)},\\
    d_4= \frac{t_2}{2i} \:(e^{ik} \cos{\theta_R} - e^{-ik} \cos{\theta_L)}.\\
\end{split}
\end{equation}

Here, for simplicity, we have set $\phi=0$.

\noindent
\textbf{Symmetries in our model.} Under general conditions, as shown in Fig.~\ref{Fig:Model}(a), our model exhibits a sublattice symmetry (SLS), satisfying $ H(k)= - S H(k) S^{-1}$, and consequently belongs to symmetry class A. In this case, the symmetry operator is $S=\sigma_z \otimes \sigma_0$ and $S^2=1$. However, with $\phi=0$, our Hamiltonian additionally respects time-reversal symmetry (TRS), with $T_+=\sigma_0 \otimes \sigma_z$ and $T_+ T_+^* = +1$, characterized by the equation $H(-k)=T_+ H^*(k) T_+^{-1}$, placing our model in the non-Hermitian symmetry class AI. This is particularly significant as it marks one of the rare models proposed in one dimension to possess this unique symmetry~\cite{kawabata2019topological}. In this symmetry class, the symmetries TRS and PHS$^\dagger$ undergo a topological unification, a feature also displayed by our model. The existence of non-trivial topological phases in symmetry class AI is exclusively observed in non-Hermitian systems, without any counterparts in Hermitian systems.\\

We compute the energy bands of our model and analytically derive the condition for EPs in order to obtain the phase transition lines. The eigenvalue equations can be computed as follows:

\begin{widetext}

\begin{gather*}
E_{1,2}(k)= \pm \sqrt{\:t_1}\:\sqrt{\:t_1 + e^{ik}\:t_2 \:(\cos{\theta_R}-\sin{\theta_R})},\\
E_{3,4}(k)= \pm \sqrt{\:e^{-ik}\:( e^{ik}\:t_1 +2\: t_2 \cos{\theta_L}) \:(\:t_1+e^{ik}\:
t_2\:(\cos{\theta_R}+\sin{\theta_R}))}.
\end{gather*}

\end{widetext}

An intriguing feature observed in our model is the exclusive dependence of the $E_1$ and $E_2$ energy bands on the parameter $\theta_R$, while the $E_3$ and $E_4$ bands are dependent on both  $\theta_L$ and $\theta_R$. This separation of parameter dependence allows for independent control and manipulation of these distinct energy bands through the corresponding spin parameters.

The EPs can now be analytically determined by computing the gap-closing points of the energy bands and verifying that the phase rigidity at that point approaches zero~\cite{rotter2009non,bulgakov2006phase}. Furthermore, by investigating the scaling behaviour of the phase rigidity on approaching an EP, we can discern the order of the EP. In our system, the phase boundaries are EP2 and EP4 contours (as depicted in Fig.~\ref{Fig:PhaseDiagram}). The phase boundary conditions are decoupled in $\theta_R$ and $\theta_L$, which is a consequence of the decoupled nature of the energy bands.  $E_1$ and $E_2$ gap closing points occur at $ \sin{\theta_R}-\cos{\theta_R}=\pm t_1/t_2$, whereas, $E_3$ and $E_4$ gap closings occur at $\cos{\theta_L}=\pm t_1/2 t_2$ and $\sin{\theta_R}+\cos{\theta_R}=\pm t_1/t_2$. We set $t_1=t_2 =1$ for subsequent calculations. The nature of the energy bands around different order EPs have been examined in detail in the supplement~\cite{supplement}.


\section{Emergence of two-fold spectral topology and flat bands}

\begin{figure} [t]
    \centering
    \includegraphics[width=0.48\textwidth]{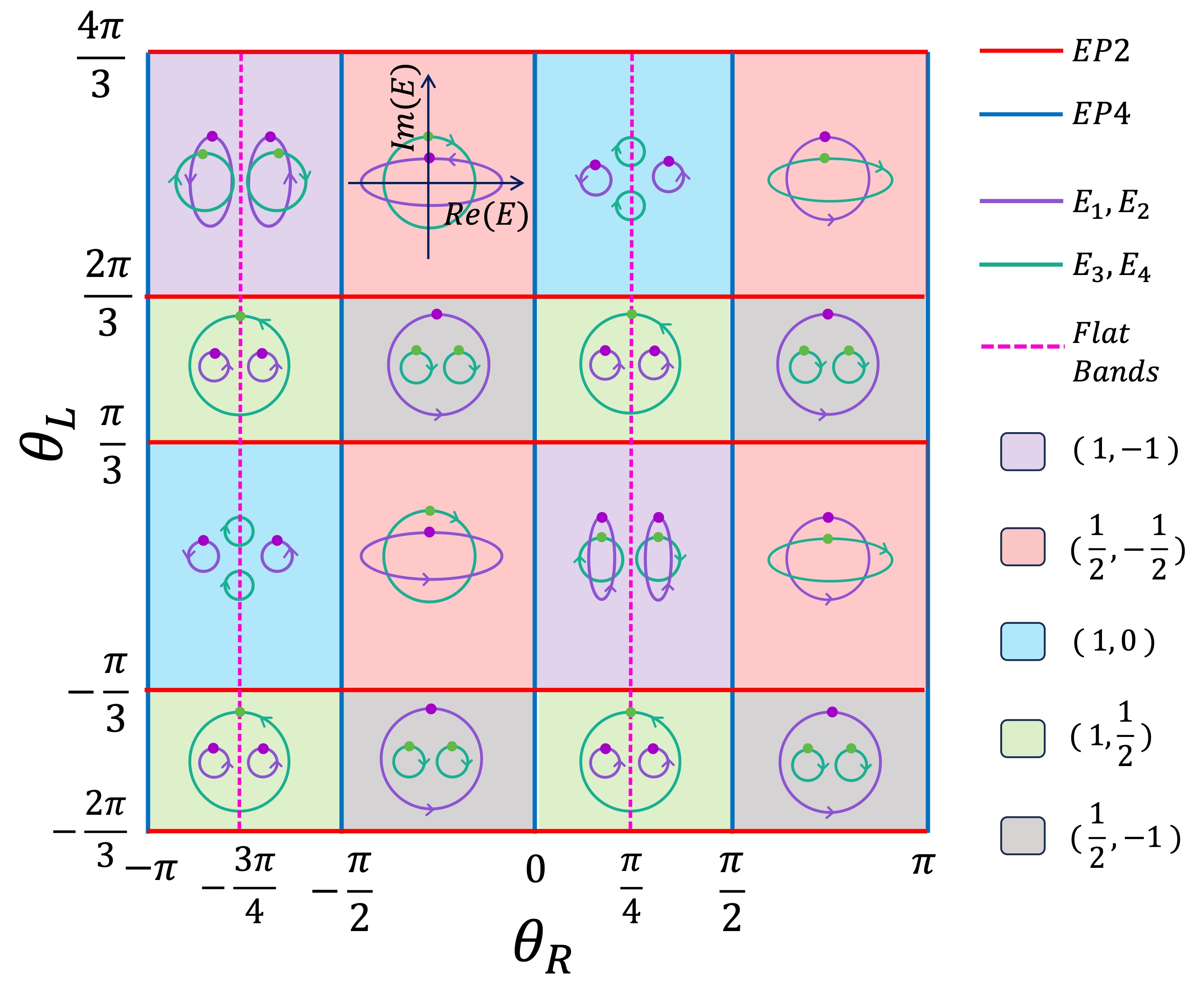}
    \caption{\label{Fig:PhaseDiagram} \textbf{Phase diagram and two-fold spectral topology.} The phase diagram shows five distinct topological phases as a function of the spin parameters, $\theta_L$ and $\theta_R$. The red and blue lines marking the phase boundaries denote EP2 and EP4 transition lines, respectively. The spectral topology associated with the $E_1, E_2$ energy bands are shown in purple in each phase, while $E_3, E_4$ spectra are shown in green. The corresponding arrows illustrate the direction of spectral winding. The associated coloured circles denote the dynamically robust chiral modes associated with each spectral loop. The paired topological invariant for each phase is presented in the legend. The pink dashed lines illustrate parameter ranges which result in two non-dispersive flat bands. Here we set $t_1=t_2=1$ and $\phi=0$.}
\end{figure}
The contours of EPs reveal the phase transition lines, partitioning the phase space $\theta_L-\theta_R$ into topologically distinct regions. In particular, from the general conditions for obtaining EPs, we find that EP4 (fourth-order EP) phase transitions occur for $\theta_R = 0,\: \pm \pi/2, \: \pm \pi$ (shown by dark blue vertical lines in Fig.~\ref{Fig:PhaseDiagram}); whereas, the EP2 (second-order EP) transitions occur along $\theta_L= \pm \pi/3,\: \pm 2\pi/3,\: \pm 4\pi/3$ (shown by red horizontal lines in Fig.~\ref{Fig:PhaseDiagram}). These exceptional contours divide the parameter space into different phases, which can be classified uniquely through the spectral topology and our newly defined topological invariant -- paired winding numbers. 

A striking feature of our non-Hermitian spin-engineered model is the emergence of a two-fold spectral topology. Spectral topology refers to the topological structure of the complex energy spectra, usually featuring distinct point-gaps and line-gaps~\cite{kawabata2019symmetry}, unique for each phase. The complex spectra are generally arc-like or loop-like, signifying the occurrence of the non-Hermitian skin effect for the latter under open boundaries~\cite{zhang2022universal}. In non-Hermitian models explored so far, each phase has one distinctive spectral topology, which is contributed by all the underlying energy bands of the system. In our model, each pair of energy bands ($E_1, E_2$) and ($E_3, E_4$) individually contribute to a different spectral topology which can be independently characterized. The spectral topology is thus rendered two-fold, necessitating a two-fold topological classification of the rich phase diagram obtained. The distinct spectral topology uncovers five different topological phases (illustrated in Fig.~\ref{Fig:PhaseDiagram}). The purple and green curves show the complex spectra contributed by ($E_1, E_2)$ and ($E_3, E_4)$, respectively. The filled circles on top of each complex loop denote the dynamically stable chiral modes. These are the energy modes which persist after a non-unitary long-time evolution of the system with lifetimes $\sim \hbar/\textup{Im}(E_\textrm{chiral})$~\cite{lee2019topological}. Under adiabatic deformations, owing to its topological origin, each spectral loop survives. Consequently, the associated chiral mode also survives, rendering them topologically robust attributes, using which we define a topological invariant $(\nu,\gamma)$ for each pair of energy bands,

\begin{equation} \label{windingno}
\begin{split}
    \nu= \frac{1}{2} \sum_{n=1,2} \text{sgn} \:[\textup{Im} \: E_n(k_{n\alpha})] \: \text{sgn}\:[\partial_k\: [\textup{Re} \: E_n(k_{n\alpha})]], \\
    \gamma= \frac{1}{2} \sum_{n=3,4} \text{sgn} \:[\textup{Im} \: E_n(k_{n\alpha})] \: \text{sgn}\:[\partial_k\: [\textup{Re} \: E_n(k_{n\alpha})]].
\end{split}
\end{equation}

Here ${k_{n\alpha}: \textup{Re}\: E_n(k_{n \alpha})= E_{b,\alpha}}$, where $E_{b,\alpha}$ denotes the set of point gaps of each spectral loop, which essentially locates the corresponding chiral mode. The first sign function takes into account the sign of the imaginary energy of the chiral mode, while the second sign function tracks the velocity of the chiral mode. The Nielsen-Ninomiya theorem~\cite{bessho2021nielsen} establishes that chiral modes always come in pairs of opposite chirality. However, the effective chirality of long-time dynamics is characterized by $\textup{Im}(E)>0$ modes~\cite{banerjee2022chiral}. Thereby, the paired winding numbers $(\nu,\gamma)$ defined above track the chirality of the dynamically robust chiral modes of each spectral loop. Each phase is uniquely classified by its characteristic paired winding number. Additionally, for a topological transition between any two phases, phase 1: $(\nu_1,\gamma_1)$ and phase 2: $(\nu_2,\gamma_2)$, we introduce $\Lambda= \abs{\abs{\nu_1}-\abs{\nu_2}}+\abs{\abs{\gamma_1}-\abs{\gamma_2}}$. If $\Lambda =1/2$, then the phase transition must cross an EP2 line, while $\Lambda=1$ denotes a phase transition across an EP4 line. This is also true for non-adjacent phases. For instance, the shortest path for transiting between the blue $(1,0)$ and grey $(1/2,-1)$ phases requires crossing one EP2 as well as one EP4 line -- here $\Lambda=3/2$. The distinct characteristic features of each phase are summarized in Table~\ref{Tab:Phases}.

\renewcommand*\arraystretch{1.5}
\begin{table}
\centering
\caption{\label{Tab:Phases} \textbf{Characterization of different phases.} The paired winding number, the number of dynamically stable chiral modes, the nature of line gap, and the skin effect characterize each phase uniquely. Here, all paired quantities refer to the spectral physics of $(E_{1,2}, E_{3,4})$. Here $+(-)$ in the winding number refers to counter-clockwise (clockwise) spectral winding. The indicative color corresponds to the representation in the phase diagram in Fig.~\ref{Fig:PhaseDiagram}. $\textup{Re/Im}$ imply the presence of a real/imaginary line gap in the respective spectra, whereas `No' refers to no line gap. The contents in the column `skin effect' indicates the direction of the skin effect due to each band-pair.}
\begin{tabular}{p{1.6cm } p{1.6cm} p{1.3cm} p{1.5cm} p{1.9cm}}
 \hline
 \hline
  Paired winding no. & Indicative colour & Chiral modes & Line gap & Skin effect \\
  \hline
  \hline
 $(1,-1)$ & Purple & (2,2) & (Re, Re) & (Left, Right) \\
 $(\frac{1}{2},-\frac{1}{2})$ & Red & (1,1) &  (No, No) & (Left, Right) \\
 $(1,0)$ & Blue & (2,0) & (Re, Im) & (Left, Right) \\
 $(1,\frac{1}{2})$ & Green & (2,1) & (Re, No) & (Left, Left) \\
 $(\frac{1}{2},-1)$ & Grey & (1,2) & (No, Re) & (Left, Right) \\

  \hline
  \hline
\end{tabular}
\end{table}

Another remarkable consequence of our model is that by solely tuning the spin parameter, $\theta_R$, we can generate two dynamically robust flat bands at $\theta_R= \pi/4,\: 3\pi/4\: (\: \forall \: \theta_L)$. $E_1$ and $E_2$ spectral loops shrink to points on the real energy axis, resulting in two flat bands at $\textup{Re}(E)=\pm t_1$ and $\textup{Im}(E)=0$. These flat bands have infinite lifetimes and are robust to non-unitary time dynamics, thereby making them accessible to experiments. Later, we further elucidate the origin of these flat bands by mapping our model to a toy model of coupled Hatano-Nelson chains. Moreover, as a consequence of the point-like spectra exhibited by $E_1,E_2$ under the flat band condition, the complex spectral area of the two left-chiral bands is lost. Consequently, the skin effect vanishes at the left edge of the lattice model. Thus, exclusively under the flat band condition, the occurrence of the skin effect is limited solely to the right edge of the lattice. We present further details in the supplementary information~\cite{supplement}.\\

As illustrated in Fig.~\ref{Fig:PhaseDiagram}, the spectral topology in each phase has point gaps. Noticeably, the line gaps observed for each spectral pair in different phases are unique to their corresponding phase. Hence, we can use the nature of line gaps as an additional tool to uniquely characterise each phase (see Tab.~\ref{Tab:Phases}). For instance, in the `purple' phase both spectral pairs have a real line gap. Contrarily, in the `blue' phase, the first spectral pair $(E_1,E_2)$ exhibits a real line gap whereas, $(E_3,E_4)$ shows an imaginary line gap. This utilization of the distinct nature of line gaps in each spectral pair enhances our ability to differentiate between the various phases of our spin-engineered SSH model. This is complemented by the quantitative characterization provided by the paired topological invariant for each phase.

To establish the robust and general nature of our discovery of the two-fold spectral topology and the two-fold topological invariant, we extend our analysis to consider the impact of next-nearest neighbour hopping interactions (refer to the supplementary materials~\cite{supplement} for detailed information). Our investigation demonstrates that, even under substantial strengths of next-nearest neighbour hopping, the two-fold spectral topology persists across all phases. This underscores the resilience and broad applicability of our discovery, solidifying its significance in diverse scenarios.

\begin{figure*} [t]
    \centering
    \includegraphics[width=0.95\textwidth]{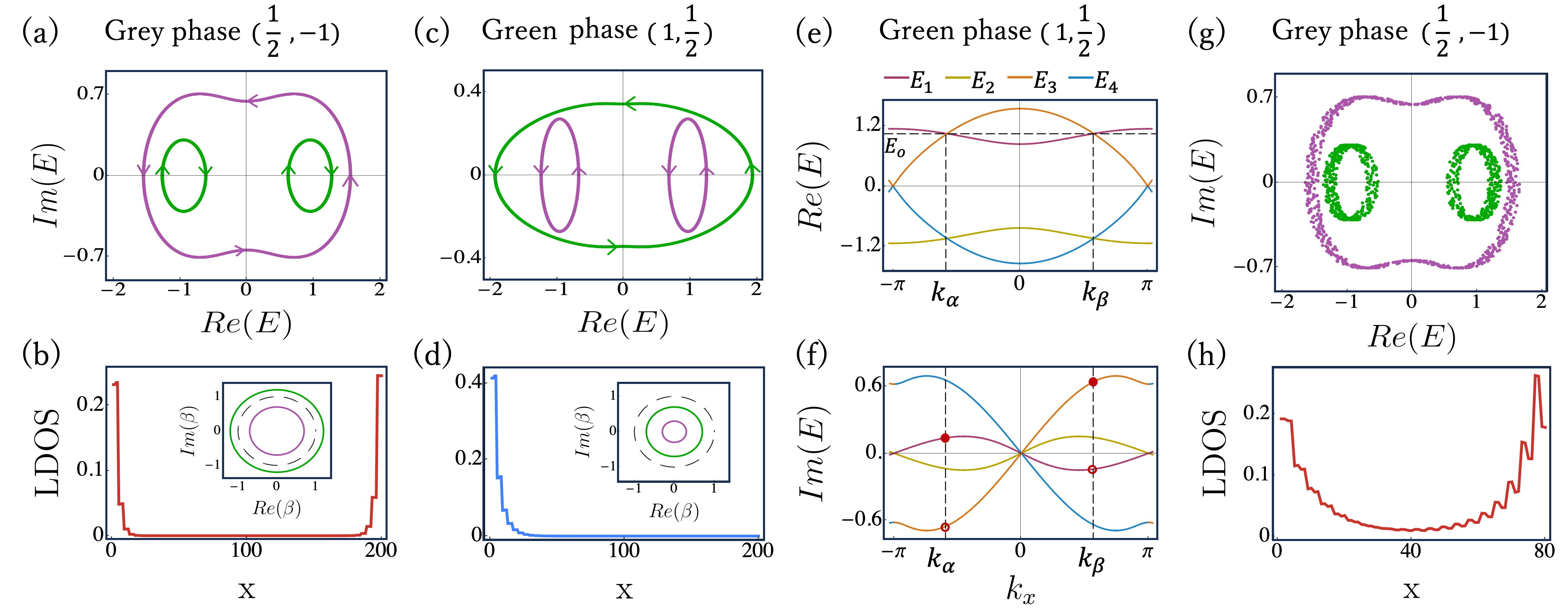}
    \caption{\label{Fig:SkinEffect} \textbf{Spectral winding, generalized Brillouin zone, and tunable skin effect.} (a) The spectral topology in the `grey' phase showing that the spectral winding of $E_1, E_2$ (in purple) is counter-clockwise while that of $E_3, E_4$ (in green) is clockwise. This results in (b) a non-Hermitian skin effect at both edges depicted by the local density of states (LDOS). The inset shows the GBZ (in corresponding colours) with $|\beta(E_1, E_2)|<1$ and $|\beta(E_3,E_4)|>1$. The black dashed lines mark the unit circle. In the green phase, (c) the direction of spectral winding is counter-clockwise for both pairs of bands, resulting in (d) skin effect only at the left edge. Here both $|\beta(E_1,E_2)|,\:|\beta(E_3,E_4)|<1$. (e)-(f) The real and imaginary spectra in the `green' phase as a function of $k_x$. A base energy $E_0$ intersects $\textup{Re}(E_1)$ and $\textup{Re}(E_3)$ at $k_\alpha$ and $k_\beta$, respectively. The corresponding $\textup{Im}(E)$ are marked in filled (unfilled) circles for long-time surviving (evanescent) modes. For $E_1$ ($E_3$) the mode at $k_\alpha$ ($k_\beta$) survives owing to higher $\textup{Im}(E)$. At $k_\alpha$, $E_1$ has a left-mover while at $k_\beta$, $E_3$ has a left-mover. Hence, skin effect in the green phase occurs at the left edge. (g)-(h) Behaviour of the spectra and skin effect under disorder ($\Delta=0.1$) in the grey phase, illustrating their robustness. Here $t_1=t_2=1, \phi=0$. For grey phase: $\theta_R= 3 \pi/4, \theta_L= 2 \pi/5$. For green phase: $\theta_R= \pi/8, \theta_L= 2 \pi/5$.}
\end{figure*}

\section{Correspondence to skin effect and chiral dynamics}
We now illustrate how our paired winding numbers ($\nu,\gamma$) can predict the skin effect under open boundary conditions. Notably, our two-fold spectral topology also renders the resultant non-Hermitian skin effect two-fold. We find that in all phases (except the `green' phase), $\:\nu$ and $\gamma$ have opposite signs, signifying that the winding direction and, thereby, the chirality of the dynamically stable modes are opposite. In contrast, in the green phase with a paired winding number (1,1/2), the chirality of the two pairs of energy bands is in the same direction. This peculiarity is reflected in the skin effect. To exemplify this, we contrast the bi-chiral `grey' phase and the uni-chiral `green' phase in Fig.~\ref{Fig:SkinEffect}(a)-(d). The purple curves denote the spectral topology of energy bands ($E_1, E_2$), while the green curves show ($E_3, E_4$). The opposite signs of the winding number in the grey phase [Fig.~\ref{Fig:SkinEffect}(a)-(b)] manifests as a `bilateral' skin effect. The counter-clockwise propagating energy bands ($E_1, E_2$) lead to skin effect at the left edge, whereas the clockwise propagating bands ($E_3, E_4$) induce skin effect at the right edge. The generalized Brillouin zone (GBZ) denoted by $\beta$ [inset of Fig.~\ref{Fig:SkinEffect}(b)] shows that $\abs{\beta(E_1, E_2)}<1$ (in purple) whereas, $\abs{\beta(E_3, E_4)}>1$ (in green) -- this establishes the directionally decoupled skin effect. In contrast, in the uni-chiral green phase [see Fig.~\ref{Fig:SkinEffect}(c)-(d)], the winding direction of both pairs of energy bands is counter-clockwise, resulting in a `unilateral' skin effect at the left edge. The GBZ shows that both $\abs{\beta(E_1, E_2)}<1$ and $\abs{\beta(E_3, E_4)}<1$. Detailed calculations of the GBZ are presented below.\\

\noindent
\textbf{Determination of the generalized Brillouin zone.} The GBZ helps establish a modified, non-Bloch bulk-boundary correspondence in a non-Hermitian system~\cite{yao2018edge,yokomizo2019non}. In our synthetic spin-designed SSH model, we outline the procedure to compute the GBZ. By analyzing the GBZ, we can determine the direction of the skin effect under open boundaries directly from the periodic system.\\
In the periodic Hamiltonian given by $H(k)$, we map $e^{ik} \rightarrow \beta$ to obtain $H(\beta)$. Here $\beta$ is, in general, a complex quantity, i.e., $\beta =\abs{\beta}\: e^{i \theta}$, where $\theta \in [0,2\pi]$. From the characteristic equation $\abs{H(\beta)-E I}=0$, we obtain the roots $\beta(E)$. Interestingly, the characteristic polynomial is of the third order, indicating that we obtain three distinct roots of $\beta$. We find that the functional dependence of the roots on the spin parameters are as follows: $\beta_1(\theta_R)$, while $\beta_{2,3} (\theta_R,\theta_L)$. This is a consequence of the decoupled nature of the energy bands, which show a similar functional dependence: $E_{1,2}(\theta_R)$, while $E_{3,4}(\theta_R, \theta_L)$. This suggests that the GBZ contributed by energy bands ($E_1, E_2$) is determined by $\beta_1$, whereas, the GBZ associated with ($E_3, E_4$) is determined by $\beta_{2,3}$. Hence, we treat the computation of the two independently.

In order to obtain the GBZ, it is necessary for the roots to satisfy the condition, $\abs{\beta_i}=\abs{\beta_{i+1}}$. By identifying the trajectory that satisfies this condition, we can determine the boundaries and the shape of the GBZ. To calculate $\beta(E_3,E_4)$, we compute the energy $E(\theta_R,\theta_L)$ for which $\abs{\beta_2}=\abs{\beta_3}$. This gives us an expression for $\beta(E_3,E_4)$, which predicts the direction of the skin effect contributed by the spectral topology of $(E_3,E_4$) correctly in every phase. When $\abs{\beta(E_3,E_4)}>1$, it signifies that there is a skin effect at the right edge. Conversely, $\abs{\beta(E_3,E_4)}<1$ implies the presence of skin effect at the left edge. The modified Brillouin zone can be expressed as follows
\begin{equation}
    \abs{\beta(E_3,E_4)}= \frac{\sqrt{2\: \cos{\theta_L}\: (\cos{\theta_R}+\sin{\theta_R})}}{\cos{\theta_R}+\sin{\theta_R}}
\end{equation}

To compute $\beta(E_1,E_2)$, we begin by identifying the energy values contributed by the ($E_1,E_2$) bands. Among these energies, we can select one and substitute it into the expression for $\beta_1$ to thereby determine the trajectory of the GBZ associated with $(E_1, E_2)$. This gives us the correct behaviour of the GBZ corresponding to every phase. This has also been presented in the Fig.~\ref{Fig:SkinEffect}. Furthermore, using the GBZ we can accurately determine the topological phase transition lines of the system under open boundary conditions~\cite{yao2018edge}. For this, we equate $\abs{\beta}=r$, where,
\begin{equation}
    \abs{\beta_{E\rightarrow0}}=r.
\end{equation}
This gives us the phase boundaries of the finite lattice system, which turns out to be consistent with the phase diagram under OBC we present in the supplement~\cite{supplement}.

Further, we highlight the physical origin of the directionally-dependent skin effect in our model in Fig.~\ref{Fig:SkinEffect}(e)-(f) for the green phase, which exhibits a unilateral skin effect. The different colours correspond to the different energy bands specified in the legend. We consider a real excitation energy $E_0$, which intersects $\textup{Re}(E_1)$ and $\textup{Re}(E_3)$ at $k_\alpha$ and $k_\beta$, respectively. At $k_\alpha$, $E_1$ has a left-moving mode while $E_3$ has a right-mover. Contrarily, at $k_\beta$, $E_1$ has a right-mover while $E_3$ has a left-mover. The net flow of eigenstates would counterbalance in the Hermitian regime. However, the inherent non-Hermiticity -- a consequence of our engineered spin choice -- associates a finite lifetime with every eigenmode, leading to a dynamically persistent current. Owing to the non-unitary time evolution $\sim e^{-iEt} \rightarrow e^{-i \mathrm{Re}(E)t} \: e^{\mathrm{Im}(E)t}$, the eigenmodes with negative imaginary energy die out over a timescale $\sim \hbar/\textup{Im}(E)$. For energy band $E_1$, the left-mover at $k_\alpha$ survives, and for $E_3$, the left-mover at $k_\beta$ survives. The persistent modes are denoted by filled red circles, and the corresponding evanescent modes are shown using unfilled red circles. Thus, the cumulative flow of states in the green phase is leftwards, leading to the unilateral skin effect at the left edge. The nature of spectral winding and skin effect in the other phases have been detailed in the supplementary materials~\cite{supplement}.

Our model facilitates control over the direction of the skin effect by suitably tuning parameters $\theta_L$ and $\theta_R$. Due to the topological origin of the skin effect and the point-gap spectra, we expect them to be robust to moderate amounts of disorder, which would make the observation of the bilateral skin effect experimentally feasible. To illustrate this robustness, we introduce random on-site disorder in our model as, $H_\textup{dis}= \sum_i \Delta_i (c^\dagger_{A,i} c_{A,i} + c^\dagger_{B,i} c_{B,i})$, where $\Delta_i= \Delta [-1,1]$ is chosen randomly and the disorder strength is $\Delta$. Fig.~\ref{Fig:SkinEffect}(g)-(h) show the robust nature of the spectral topology and bilateral skin effect in the exemplary grey phase at disorder strength $\Delta=0.1 \:t_1$, confirming our expectations.\\


\begin{figure} [t]
    \centering
    \includegraphics[width=0.45\textwidth]{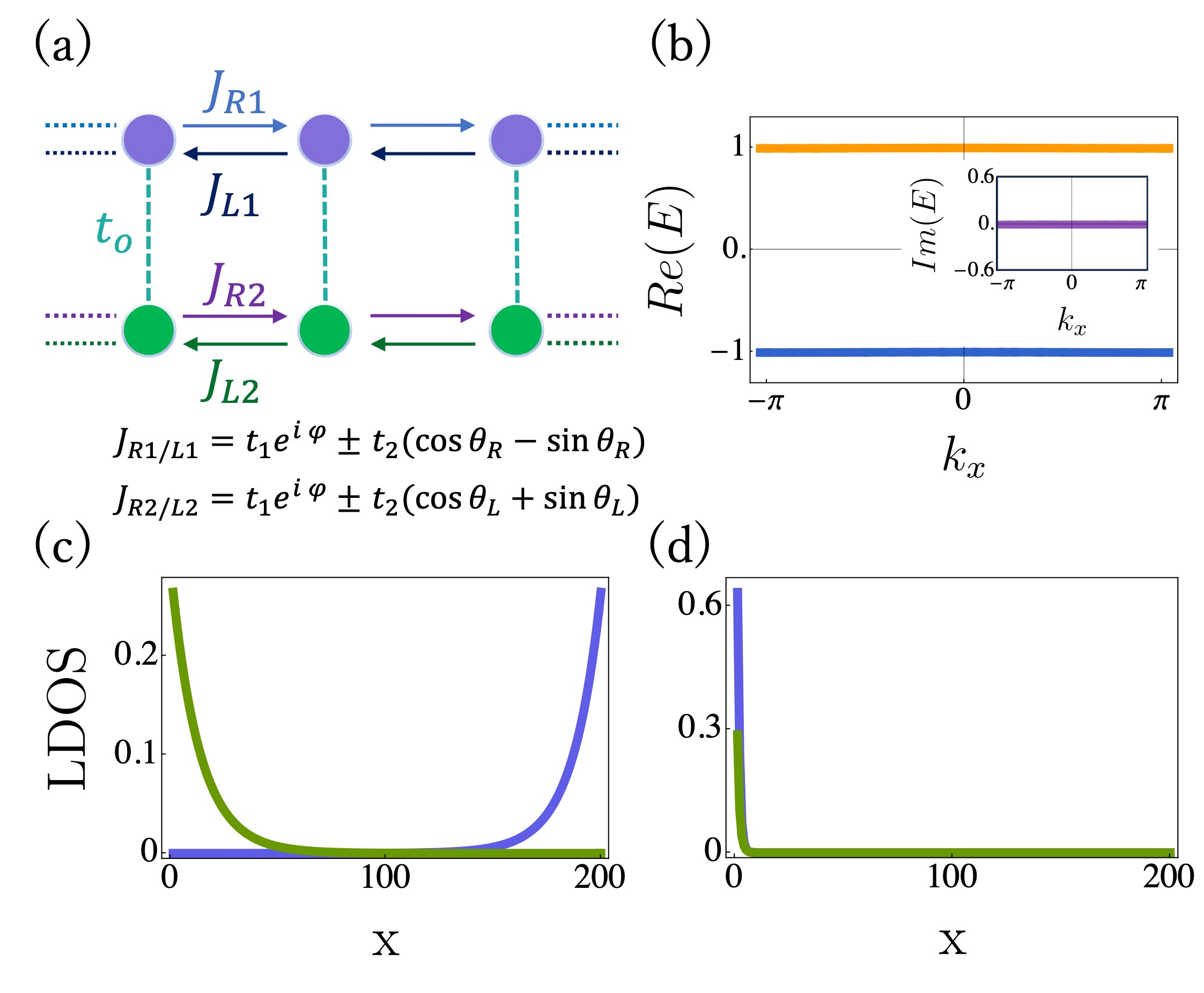}
    \caption{\label{Fig:HatanoNelson} \textbf{Coupled Hatano-Nelson chains as toy model.} (a) Two Hatano-Nelson chains linked using coupling strength $t_o$. The hopping parameters are denoted below the schematic, mapping it to our spin-engineered SSH model. (b) Dynamically robust ($\textup{Im}(E)=0$) flat bands at $\textup{Re}(E)=\pm t_o$, in the strong coupling limit ($t_1/t_o=0.01$). (c) Bilateral skin effect and (d) unilateral skin effect attained by tuning parameters $\theta_L$ and $\theta_R$ in the weak coupling limit ($t_o/t_1=0.1$). Here $t_2=1,\phi=0$ for all plots. We set (b) $\theta_{R,L}= \pi/4, 3\pi/4$\:, (c)  $\theta_{R,L}= \pi/2, \pi/2$\:, and (d)  $\theta_{R,L}= -\pi/5, 3\pi/4$.}
\end{figure}

\begin{figure*} [t]
    \centering
    \includegraphics[width=0.75\textwidth]{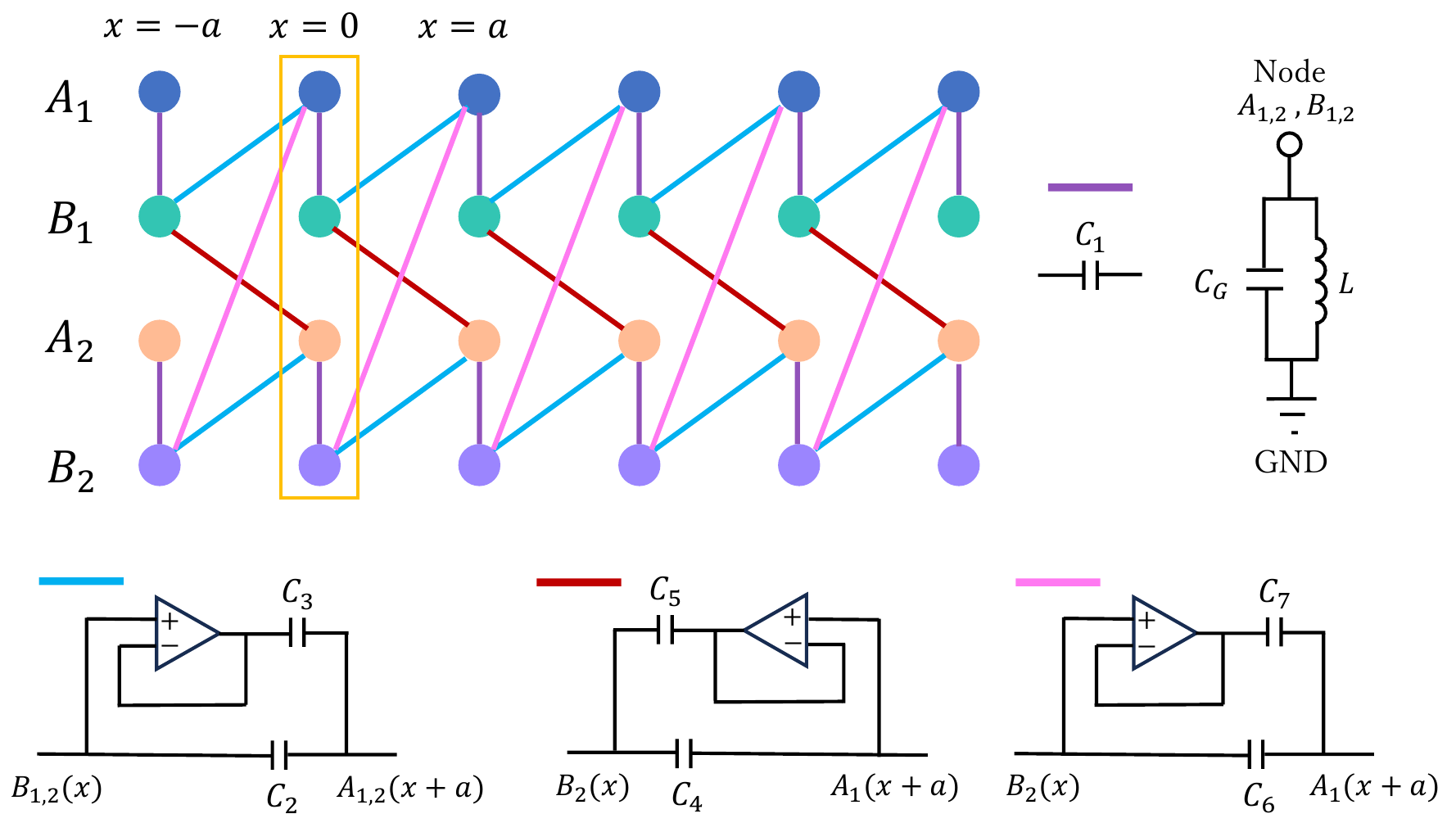}
    \caption{\label{Fig:Topoelectric} \textbf{Topoelectrical circuit to realize the spin-engineered SSH model.} The four unit cell sites of our model, $A_1, A_2, B_1$, and $B_2$, form a chain of nodes in the topoelectrical circuit. Each node is grounded through the $C_G-L$ circuit shown in the top right panel. The connections between the nodes are denoted by the different coloured lines, each specified in the legend. The non-reciprocal nature of hopping is explicitly obtained using directional amplifiers, whereas the Hermitian hopping terms are achieved through a purely capacitive connection.}
\end{figure*}

\section{Mapping to a toy model: coupled Hatano-Nelson chains}
To elucidate the controllable nature of our proposed direction-selective skin effect, we devise a simple toy model. We consider coupled chains of the Hatano-Nelson model, which is the simplest non-Hermitian model with rich topological features~\cite{hatano1996localization}. As illustrated in Fig.~\ref{Fig:HatanoNelson}, we couple two Hatano-Nelson chains with non-reciprocal hopping, with coupling strength $t_o$. The hopping parameters are set as represented in the schematic, in order to map it to our original spin-engineered SSH model. In the weak coupling limit, i.e., $t_o/t_1<<1$, we discover that by tuning parameters $\theta_L$ and $\theta_R$, we can control the direction of the skin effect. Remarkably, we find that it is possible to obtain the bilateral skin effect [Fig.~\ref{Fig:HatanoNelson}(c)], as well as the unilateral skin effect [Fig.~\ref{Fig:HatanoNelson}(d)], even for fixed values of $t_1$ and $t_2$. Furthermore, in the strong coupling limit, i.e., $t_o/t_1>>1$, we can tune the free parameters such that $\theta_R=\pi/4$ and $\theta_L=3\pi/4$, leading to the emergence of two non-dispersive flat bands at $\textup{Re}(E)=\pm t_o$. Similar to our original model, these flat bands are also dynamically robust with $\textup{Im}(E)=0$ [shown in Fig.~\ref{Fig:HatanoNelson}(b)]. In the strong coupling limit, there is a dissociation process, resulting in unit cells that consist of one atom each from the upper and lower chains. In this regime, the wavefunction within each unit cell becomes localized -- this leads to the emergence of a flat band, with the wavefunction pinned to its respective unit cell.\\

\section{Experimental Realization: Topoelectrical Circuit}
To realize the two-fold spectral topology in our non-Hermitian spin-engineered SSH model, we propose a one-dimensional topoelectrical circuit lattice consisting of capacitors, inductors, and operational amplifiers~\cite{goren2018topological,luo2018nodal,imhof2018topolectrical,zhang2020non,zhang2021tidal,xu2021coexistence}. The operational amplifier plays a crucial role as the fundamental component of the circuit cell, introducing non-Hermiticity to the system. The differential inputs of the operational amplifier are characterized by a non-inverting input $(+)$ with voltage $V_{+}$ and an inverting input $(-)$ with voltage $V_{-}$. Under ideal operating conditions, the output voltage $(V_{\text{out}})$ is related to the inputs as $V_{\text{out}} = A(V_{+} - V_{-})$, where $A$ represents the open-loop gain, typically significantly high for an operational amplifier. We design the operational amplifier as a voltage follower by connecting the inverting input $(-)$ to the output, resulting in $V_{-}=V_{\text{out}}$. It corresponds to an output voltage $V_{\text{out}}=A(V_{+}-V_{\text{out}}) \implies V_{\text{out}} =\frac{A}{A+1}V_{+} \approx V_{+}$. Consequently, the input terminals $V_{+}$ and $V_{-}$ satisfy the virtual short circuit criterion of an ideal operational amplifier. By implementing Kirchoff's current law we design a circuit lattice corresponding to the tight-binding model as shown in Fig.~\ref{Fig:Topoelectric}. The corresponding current equation at every node can be found in the supplement~\cite{supplement}.

Consequently, the Kirchhoff equations can be cast as an eigenvalue-like equation (assuming lattice spacing $a$ to be 1), $I =i \: M(k) V$, where $I$ and $V$ are vectors of the form: $I=[I_{A_1},I_{A_2},I_{B_1},I_{B_2}]$ and $V=[V_{A_1},V_{A_2},V_{B_1},V_{B_2}]$.
The exact form of matrix $M(k)$ can be found in the supplement~\cite{supplement}.
By performing a straightforward mapping, we can associate our tight-binding Hamiltonian with the circuit lattice. Hereby, the hopping parameters align with the capacitors, the wave functions correspond to the voltage, and the eigenvalues relate to the resonance frequency. We derive the parameters to realize our original model which are summarized as follows: $t_1=\omega C_1, \:\: t_2 \cos{\theta_L}=\omega C_2,\:\: -it_2\cos{\theta_L}=\omega C_6, \:\: it_2\cos{\theta_L}=\omega C_4, \:\: t_2 \cos{\theta_R}=\omega (C_2+C_3), \:\:-it_2\sin{\theta_R}=\omega (C_4+C_5), \:\: t_2 \cos{\theta_R}=\omega (C_2+C_3)$, and $it_2\sin{\theta_R}=\omega (C_6+C_7)$. Therefore, by judiciously tuning the values of capacitors we can realize the two-fold spectral topology in the proposed one-dimensional non-Hermitian electrical circuit lattice. This allows for the manipulation of non-Hermitian spin engineering effects by analyzing the admittance spectra~\cite{dong2021topolectric,lee2018topolectrical}.\\

\section{Conclusions}
In conclusion, our work introduces a non-Hermitian spin engineering, as a means to achieve multi-fold spectral topology, particularly within the non-Hermitian symmetry class AI. We demonstrated that a two-band model can exhibit a two-fold spectral topology which in turn leads to a multi-chiral non-Hermitian skin effect. We introduced a paired winding number which characterizes the distinct spectral topology of each phase and provides insights into the nature of phase transitions through EPs, while also capturing the chiral dynamics which determines the direction of skin effect. This multi-fold spectral topology offers the possibility of directionally-controllable skin effects and mode selection. Importantly, our proposed models and their associated effects are experimentally accessible, providing a promising avenue for future studies and potential applications. As an illustration of the experimental feasibility, we have developed a topoelectric circuit that can faithfully emulate the underlying spin-engineered SSH Hamiltonian. This circuit provides a viable platform for realizing our predicted effects.

\subsection{Acknowledgments}
\noindent R.S. and A.B. are supported by the Prime Minister's Research Fellowship (PMRF). A.N. acknowledges support from the Indian Institute of Science.

\bibliography{Ref}

\end{document}